\documentclass[%
reprint,
 amsmath,amssymb,
 aps,
]{revtex4-1}

\usepackage[dvipsnames]{xcolor}
\usepackage{graphicx}
\usepackage{dcolumn}
\usepackage{bm}
\usepackage{color}
\usepackage{hyperref}
\usepackage{subfigure}
\usepackage{physics}

\newcounter{Cequ}
\newenvironment{Cequation}
   {\stepcounter{Cequ}%
     \addtocounter{equation}{-1}%
     \equation}
   {\endequation}

\begin{document}


\title{Topological soliton-polaritons in 1D systems of light and fermionic matter}

\author{Kieran A. Fraser and Francesco Piazza$^*$}
 
\affiliation{Max  Planck  Institute  for  the  Physics  of  Complex  Systems,
N\"{o}thnitzer  Stra{\ss}e  38,  D-01187  Dresden,  Germany}

\date{\today}

\begin{abstract}
  Quantum nonlinear optics is a quickly growing field with large technological promise, at the same time involving complex and novel many-body phenomena. In the usual scenario, optical nonlinearities originate from the interactions between polaritons, which are hybrid quasi-particles mixing matter and light degrees of freedom. Here we introduce a type of polariton which is intrinsically nonlinear and emerges as the natural quasi-particle in presence quantum degenerate fermionic matter. It is a composite object made of a fermion trapped inside an optical soliton forming a topological defect in a spontaneously formed crystalline structure.
Each of these soliton-polaritons carries a $\textbf{Z}_2$ topological quantum number, as they create a domain wall between two crystalline regions with opposite dimerization so that the fermion is trapped in an interphase state. These composite objects are formally equivalent to those appearing in the Su-Schrieffer-Heeger (SSH) model for electrons coupled to lattice phonons. 
\end{abstract}


\maketitle

\textbf{Introduction}

Hybrid systems involving photons and neutral atomic gases have emerged as ideal platforms for nonlinear quantum optics \cite{chang_vuletic_lukin_2014}, characterized by effective interactions between photons even at very low light intensities. This realm of optics offers interesting possibilities for quantum technologies, as optical nonlinearities at the single-photon level would facilitate quantum information processing with light, the latter being at the same time an ideal information carrier \cite{quantum_internet}. Moreover, an ensemble of photons and atoms interacting at the level of single quanta is an intriguing many-body system which has recently attracted a lot of interest \cite{chang_vuletic_lukin_2014,cavity_rmp}.

In such systems the low-lying excitations are polaritons, which emerge as long-lived quasi-particles. In the usual scenario, polaritons are a linear superposition of non-interacting photonic and atomic excitations, while the optical nonlinearities of the system are generated by the interactions between polaritons inherited from the atoms. 

Here we show that a different situation can arise, where a new type of long-lived hybrid quasi-particle emerges. This type of polariton is not a linear superposition of non-interacting atomic and photonic degrees of freedom, but rather a result of the optical nonlinearity of the system in presence of an atomic Fermi surface. It is a composite object made of a fermion trapped inside an optical soliton forming a topological defect in an emergent crystalline structure.

\textbf{Results}\newline
\textbf{System and phenomenology.}
\begin{figure}[]
\centering
\includegraphics [width=0.48\textwidth]{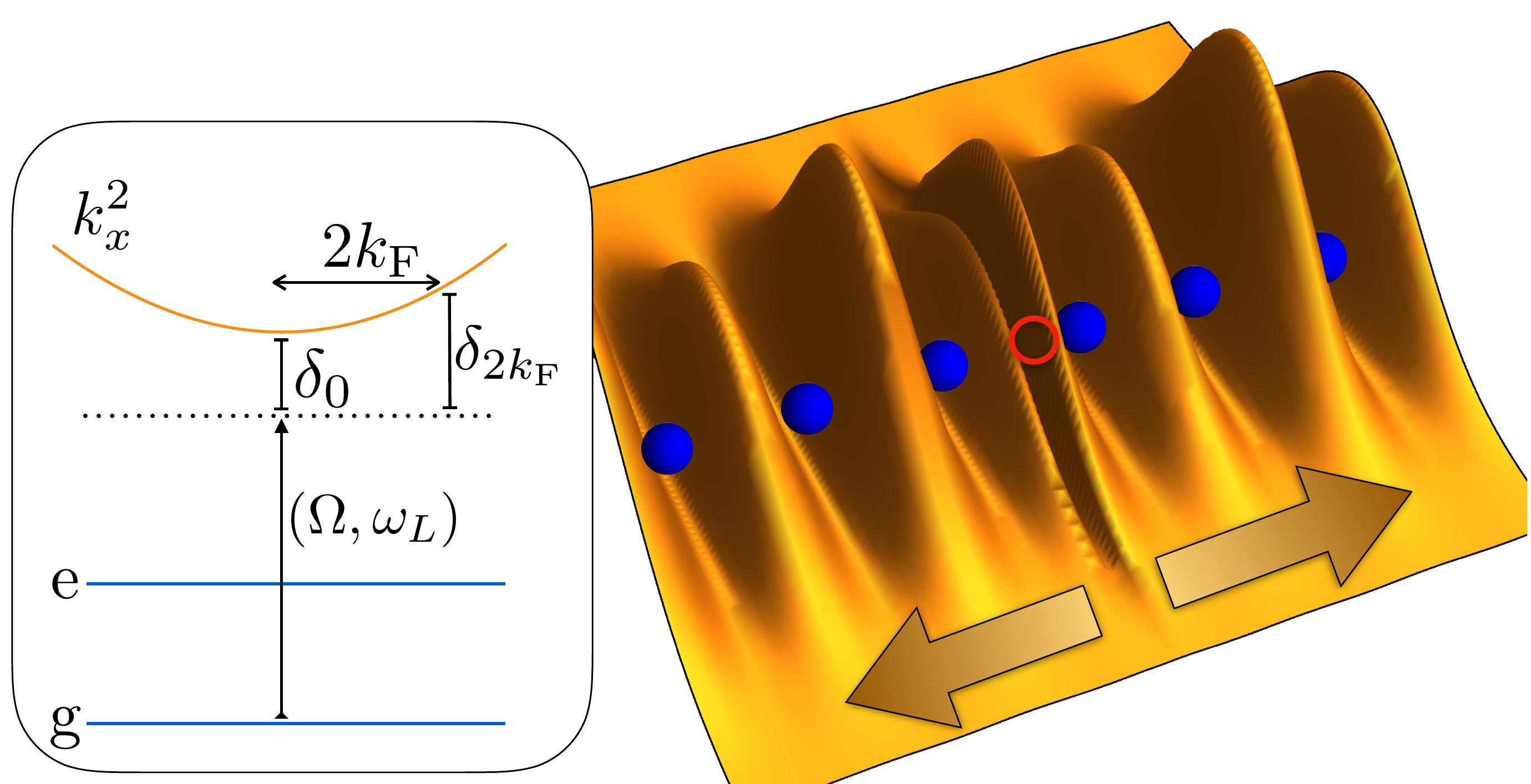}
\caption{Pictorial representation of the system. A (quasi-) one-dimensional cloud of fermionic atoms (blue spheres) is transversally driven by a coherent laser and is coupled to a set of guided modes. The corresponding light intensity (including the homogeneous background) is shown as a dark-yellow surface with a transverse exponential damping indicating the guided nature of waveguide modes (arrows indicate the propagation direction of the guided modes). The situation shown corresponds to the presence of a light-matter defect at the center where an optical soliton traps an excess particle (red circle). The optical soliton connects two spatially ordered regions with different dimerization. The inset shows the atom internal level structure (spacings not to scale) together with the dispersion of the guided electromagnetic modes. The laser frequency is $\omega_{\rm L}$, and its Rabi frequency is denoted by $\Omega$.}
\label{fig:setup}
\end{figure}
Such a crystalline structure appears when a degenerate cloud of fermions is coupled to the propagating modes of an optical waveguide in the configuration of Fig.~\ref{fig:setup} (crystallization in optical waveguides has been studied only for classical particles so far \cite{chang2013self,griesser2013light}).
In this (quasi-)one-dimensional configuration, the fermionic cloud is unstable towards density modulations with wavenumber equal to twice the Fermi momentum $k_{\rm F}$ -- analogous to the Peierls instability known since the 1950's in the context of solids \cite{peierls1955quantum} -- so that photon-mediated Umklapp scattering of atoms between the Fermi points (momenta of $\pm k_{\rm F}$) induces crystallization. This scenario has been considered for fermionic atoms coupled to a single standing-wave mode of an optical resonator \cite{keeling_fermi_2014,piazza_fermi_2014,zhai_fermi_2014,piazza_topol}.
The effect of a sharp Fermi surface is however much more prominent in the presence of multiple electromagnetic modes, as is the case in a confocal cavity \cite{gopalakrishnan2009emergent} or for the continuum of propagating modes of an optical waveguide that we consider here. Firstly, provided that $2k_{\rm F}$ is included in the wavenumbers of the waveguide's electromagnetic modes, the instability toward crystallization will be always present even at vanishing coupling, differently from the single-resonator-mode case. Secondly, spatially-ordered fermionic patterns coupled to multimode fields are sensitive to commensurability effects, and the resulting structures can accommodate hybrid light-matter defects as low-lying excitations.
As we show here, the latter consist of a fermion trapped in an interphase state located at a solitonic deformation of the electromagnetic potential. The optical soliton separates two regions of the crystal with opposite dimerization and therefore carries a $\mathbf{Z}_2$ topological quantum number. Its size is set by the inverse Fermi momentum. As anticipated above, these light-matter defects constitute a novel type of polariton which exists only in presence of optical nonlinearities. Indeed, in order to create this type of excitation, the fermion has to be transferred to the interphase state, which itself can only be created via the formation of an optical soliton i.e. an intrinsically nonlinear process.

In a specific parameter regime that we identify, our system is formally described by the continuum limit of lattice electron-phonon models, more specifically the Su-Schrieffer-Heeger (SSH) model \cite{ssh_1979,maki_1980,brazovskii1980self,rice_mele_1982}.  
The combination of its simplicity and its hosting of topologically-protected states has as of late inspired much interest in the SSH model. There have been a number of synthetic implementations in a variety of different systems, such as cold atoms in optical lattices  \cite{Bloch,Weitz,Gadway}, chains of Rydberg atoms \cite{Browaeys}, semiconductors \cite{Amo}, granular particles \cite{Yang}, resistor-inductor-capacitor (RLC) circuits \cite{Thomale} and microring resonators \cite{khaj}. 
As opposed to the above implementations, the system we consider here features an additional ingredient which is crucial for the emergence of topological defects as the low-lying excitations, namely a dynamical lattice which is self-consistently modified by the particles, as in the original SSH model (see also a recent proposal with atoms and two-level systems \cite{PhysRevLett.121.090402}).
Beyond a standard quantum-nonlinear-optics perspective, the setup proposed here thus also offers a means to further explore the interplay between topology and interactions in a controlled hybrid setup involving light and matter degrees of freedom.

\textbf{Model.}
We consider a degenerate Fermi gas of $N$ laser-driven neutral atoms interacting with the multimode radiation field of an optical waveguide in the configuration of Fig.~\ref{fig:setup}. 
The internal atomic transition between the ground state manifold to the excited electronic state is driven by a pump laser of Rabi frequency $\Omega$ and frequency $\omega_{\rm L}$, and coupled with rate $g_k$ with the waveguide's electromagnetic-field modes. The latter we separate into a series of running guided modes denoted by $\eta_{k}=e^{ikx}$. We work in the regime of large atomic detuning $\delta_{\rm A}=\omega_{\rm L}-\omega_{\rm A}$ where the population of the excited state is negligible and spontaneous emission is suppressed, so that the excited atomic state can be adiabatically eliminated. Using the rotating-wave and dipole approximations the Hamiltonian in the frame rotating at the pump frequency reads
\begin{equation}\label{Hamil}
\begin{split}
\hat{H}=&-\sum_{k}\delta_{k}\hat{a}_{k}^{\dagger}\hat{a}_{k}\\
&+\sum_{\sigma=\{\uparrow,\downarrow\}}\int dx\hat{\psi}_{\sigma}^{\dagger}(x)\bigg(-\frac{\hbar^2}{2m}\frac{d^2}{dx^2}+\hat{V}(x)\bigg)\hat{\psi}_{\sigma}(x),
\end{split}
\end{equation}
\begin{equation}
\begin{split}
\hat{V}(x)=&\sum_{k}\frac{\Omega g_{k}}{\delta_{\rm A}}\eta_{\Omega}(x)(\eta^{*}_{k}(x)\hat{a}_{k}^{\dagger}+\eta_{k}(x)\hat{a}_{k})\\
&+ \sum_{k,k'}\frac{g_{k}g_{k'}}{\delta_{\rm A}}\eta^{*}_{k}(x)\eta_{k'}(x)\hat{a}_{k}^{\dagger}\hat{a}_{k'}.
\end{split}
\end{equation}
The spin degree of freedom for the fermions indexed by $\sigma$ can be introduced using hyperfine levels within the ground-state manifold.
We assume an external trapping potential which restricts the atomic motion to along the waveguide axis, $x$, such that the momentum transfer via photon scattering between atoms is due entirely to the photons propagating along this axis. Correspondingly, we take the pump spatial mode function $\eta_{\Omega}=1$ i.e constant in space over the transverse extension of the atomic cloud. The spatially homogeneous Stark shift resulting from the pump, as well as the homogeneous component of the light scattered from the atoms, can be absorbed in the chemical potential.
We approximate the dispersion of the guided modes to be quadratic so that $\omega_{k}=\omega_0+w k^2$.
The ground-state fermionic annihilation operator is labelled $\hat{\psi}$ and satisfies the canonical anticommutation relation $\{\hat{\psi}(\textbf{r}),\hat{\psi}^{\dagger}(\textbf{r}')\}=\delta(\textbf{r}-\textbf{r}')$. The detuning of the pump from the waveguide field mode of wavenumber $k$ is denoted $\delta_{k}=\omega_{\rm L}-\omega_k$. Since the dipole coupling $g$ scales like $1/\sqrt{L}$ with $L$ the length of the waveguide, it is convenient to define couplings which are intensive in the thermodynamic limit $N,L\rightarrow\infty$, $N/L=n=$const.: $\lambda_k=\Omega g_k \sqrt{N}/\Delta_{\rm A}$ and $U=g_k^2N/\Delta_{\rm A}$. The nonlinear coupling $U$ can be arbitrarily suppressed by choosing $\Omega\gg g_k$ and will be neglected in what follows.

In order to compute the phase-diagram, low-lying excitations and optical response, we
formulate the problem within a path-integral formalism \cite{PIAZZA2013135,piazza_fermi_2014,piazza_topol,lang2017collective}, as described in the Supplementary Material, Note 1.

\begin{figure}
\includegraphics[width=0.48\textwidth]{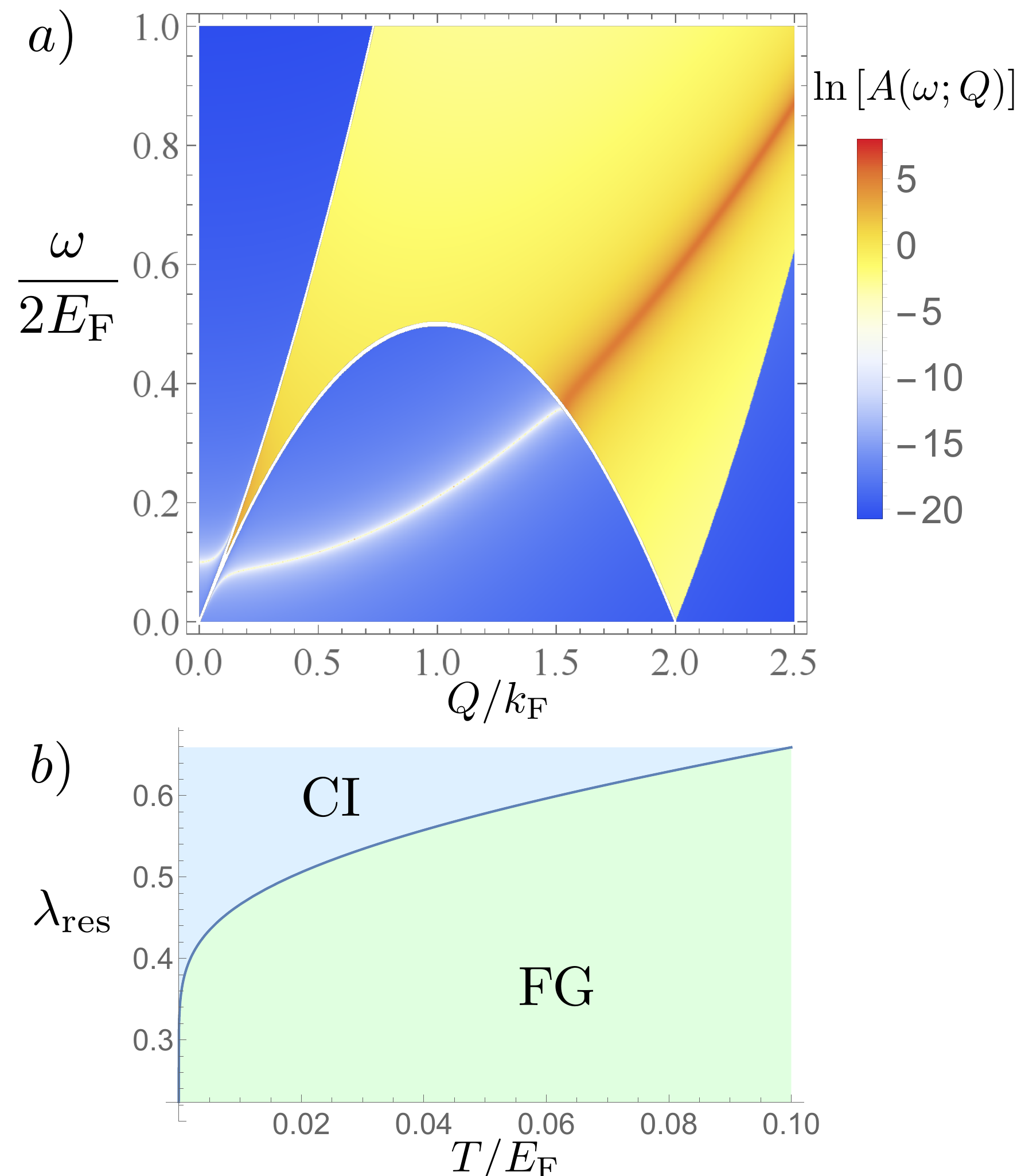}
\caption{Spectral response and phase diagram of the system. Panel a): spectral function $A(\omega;Q)$ of the photons in the spatially homogeneous phase as a function of rescaled frequency $\omega/2 E_{\rm F}$ and momentum $Q/k_{\rm F}$. The parameters are $\lambda=0.2$, $\omega_k-\omega_{\rm L}=0.1+0.125k^2$. Note the presence of the photon-like polariton peak cutting the frequency axis at $\omega/2E_{\rm F}=0.1$ and the atom-like polariton breaking off from the particle-hole continuum as a separate, well-defined mode which then rejoins the continuum and begins to broaden. Panel b): phase diagram of the system as a function of the rescaled temperature $T/E_{\rm F}$ and coupling strength $\lambda_{\rm{res}}=\lambda(2k_{\rm F})/\sqrt{-2\nu_{k_{\rm F}}/n\delta_{2k_{\rm F}}}$. At $T=0$ the critical pump strength vanishes and the system undergoes a phase transition from the Fermi gas (FG) to the crystalline insulator (CI) phase even at vanishing coupling.
}
\label{Spectral}
\end{figure}

\textbf{Crystallization into an insulator.} In Fig. \ref{Spectral}a) we show the spectral function $A(\omega,k)=-2\mathrm{Im}\int dt\exp(-i\omega t)i\theta(t)\langle [\hat{a}_k(t), \hat{a}_k^\dag(0)]\rangle$ of the system in the spatially homogeneous phase (see Supplementary Note 2).
The spectral function features an almost constant value within a region corresponding to the particle-hole continuum characterizing the 1D Fermi gas. In addition, two sharp branches are present, indicating the existence of two long-lived polaritonic quasiparticles: a photon-like polariton branch with a renormalized waveguide dispersion, and an atom-like polariton growing out of the fermionic particle-hole continuum. Differently from the topological soliton-polaritons which will appear in the crystalline phase that we describe next, the polaritons in the spatially homogeneous phase are linear excitations visible in the spectral function $A(\omega,k)$, which indeed characterizes the linear response of the system. 
At zero temperature, the spectral weight from the continuum around twice the Fermi momentum is finite down to zero frequency, where the density response diverges, and the ultracold gas is rendered unstable to density modulations corresponding to $Q=2k_{\rm F}$. In fact, the critical coupling above which a given momentum component $Q$ becomes unstable reads (see Supplementary Note 2) ${\lambda}_{\rm c}^2(Q)=-n\delta_Q/2\Pi(0; Q)$, where $\Pi(\omega; Q)$ is the dielectric (or Lindhard) function describing the density-response of the Fermi gas. The latter diverges logarithmically at zero temperature in 1D for $Q\rightarrow 2 k_{\rm F}=\pi n$ due to resonant Umklapp scattering between $\pm k_{\rm F}$. The critical coupling takes indeed the form $\lambda_{\rm c} (Q)\sim 1/\ln|1-Q/2k_{\rm F}|^{-1}$. The resulting phase diagram of the system is shown in Fig.~\ref{Spectral}b).

\begin{figure*}[htp]
\centering

\hspace{-.5cm}
\includegraphics[width=0.98\textwidth]{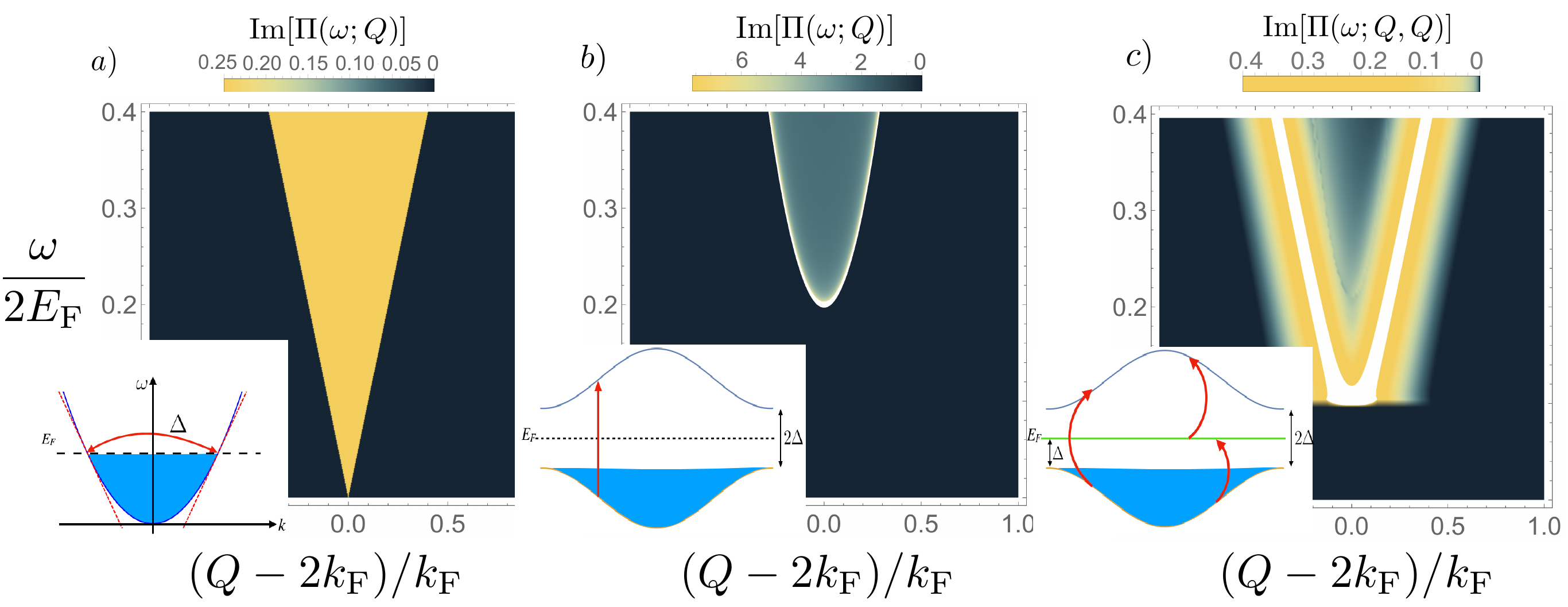}
\caption{Optical absorption spectrum in each of the three phases, given by the imaginary part of the dielectric function $\Pi(\omega,Q,Q')$, as a function of rescaled frequency $\omega/2 E_{\rm F}$ and momentum $(Q-2k_{\rm F})/k_{\rm F}$. The insets show the corresponding single-particle spectrum of the fermions. Panel a) corresponds to the spatially homogeneous phase for $\Delta_0=0.1$, which displays an ungapped absorption spectrum. Correspondingly, the particles at the Fermi points are scattered from one side of the Fermi sea to the other via Umklapp scattering of photons of wavenumber $2k_{\rm F}$. For the crystalline phase, $b$) shows a gap opening in the absorption spectrum equal to $2\Delta_0=0.2$, which can be understood from the inset figure showing a band insulator spectrum.   In the presence of the defect the gap in the spectrum is halved as seen in $c$), where the allowed optical transitions are now from the lower band to the bound state upper band, and from the bound state to the upper band. 
In c) we rescaled $\Pi(\omega,Q,Q)$ by the momentum grid spacing in order for it to have the same dimensions as in a),b).}
\label{fig:figure3}

\end{figure*}

An analogous instability was described by Peierls for electrons coupled to lattice phonons \cite{peierls1955quantum}. Differently from the Peierls instability in lattice models, where the lattice constant of an already-present lattice is doubled, the instability of our homogeneous system (where the relevant fermionic excitations are right(left) movers as shown in the inset of Fig.~\ref{fig:figure3}a)) towards a $2k_{\rm F}$-modulation spontaneously breaks the translation invariance exhibited by the continuous $U(1)$-symmetry of the Hamiltonian (\ref{Hamil}): $x\to x+a$ with $a$ a real constant. For $U=0$ there is an additional symmetry of our theory emerging at low energies, i.e. in the vicinity of the Fermi surface (see next section). It is a $\textbf{Z}_2$ symmetry involving a particle-hole $\hat{\psi}(x)\rightarrow\hat{\psi}(x)^\dag$ plus a parity $\hat{a}(x)\rightarrow-\hat{a}(x)$ transformation, with $\hat{a}(x)=\sum_k\eta_k(x)\hat{a}_k$. This symmetry plays a crucial role in the spatially ordered phase. Indeed, in breaking the continuous spatial-translation-invariance the system additionally breaks the above $\textbf{Z}_2$ symmetry by choosing between one of two possible dimerizations i.e. by choosing either the odd or the even sites of the $2k_{\rm F}$-optical lattice. The two dimerizations in this system are directly analogous to those in the SSH model.
The spatially ordered phase is an insulator since the $2k_{\rm F}$-modulation gaps out the Fermi surface (see the inset of Fig.~\ref{fig:figure3}$b$)).

The finite energy cost for exciting fermions into the conduction band leaves space for the appearance of lower-lying excitations, the latter necessarily involving lattice distortions i.e. being of polaritonic nature. As we show below, such distortions correspond to solitonic defects localized between two different $\textbf{Z}_2$-dimerizations of the optical lattice which traps a fermion in the resulting bound state (see Fig.~\ref{fig:setup}). The solitons thus have to carry either one of two topological quantum numbers related to the $\textbf{Z}_2$ symmetry. The results we present in what follows are obtained within a low-energy description of our system, which we introduce next.

\textbf{Low-energy theory.} 
The low-energy theory is obtained by considering atomic degrees of freedom only in the vicinity of the Fermi surface, thereby linearizing the dispersion around $\pm k_{\rm F}$. Correspondingly, the electromagnetic field must be restricted to momentum components around $Q=\pm 2k_{\rm F}$, i.e. those modes responsible for Umklapp scattering.
With these restrictions the effective low-energy action in the Matsubara imaginary-time formalism \cite{piazza_fermi_2014, PIAZZA2013135} becomes
$S=S_{\mathrm{a}}+S_{\mathrm{ph}}+S_{\mathrm{a/ph}}$ (see Supplementary Note 3) where:
\begin{equation}
\begin{split}
&S_{\mathrm{a}}=\int\mathrm{d}x\int\mathrm{d}\tau\sum_{\sigma=\uparrow,\downarrow}\Psi_{\sigma}^{\dagger}(\partial_{\tau}-iv_{\rm F}\partial_x\sigma_3){\Psi_{\sigma}};\\
&S_{\mathrm{ph}}=\int\mathrm{d}x\int\mathrm{d}\tau\Delta^*\bigg(\frac{\sqrt{n}}{2\lambda}\partial_{\tau}+\tilde{\delta}_{2k_{\rm F}}\bigg)\Delta;\\
&S_{\mathrm{a/ph}}=\int\mathrm{d}x\int\mathrm{d}\tau\sum_{\sigma=\uparrow,\downarrow}\Psi_{\sigma}^{\dagger}\sigma_1\mathrm{diag}(\Delta^*,\Delta){\Psi_{\sigma}}
\end{split}
\label{eq:lowenergytheory}
\end{equation}
are the atomic, photonic and interaction elements of the action, respectively and we have introduced the spinor ${\Psi(x)}=(u(x),v(x))^T$ with $u(v)$ denoting a fermionic right(left) mover: $u_k(v_k)=\psi_{k+k_{\rm F}}(\psi_{k-k_{\rm F}})$. 
These are coupled by the electromagnetic field components which are expanded in momentum space about $Q=2k_{\rm F}$ so that $\Delta_q$ corresponds to the value of the photon field at momentum $q$ relative to $2k_{\rm F}$: $\Delta_q=a_{2k_{\rm F}+q}$. At low energy we have assumed $\lambda_k\simeq \lambda$ and rescaled $(2\lambda/\sqrt{n})\Delta\to\Delta$ such that this dimerization field has units of energy. Correspondingly $\tilde{\delta}_{2k_{\rm F}}=-n\delta_{2k_{\rm F}}/2\lambda^2$, where $\delta_{2k_{\rm F}}$ is the detuning of the laser from the waveguide mode of wavenumber $2k_{\rm F}$. 

The spatially homogeneous phase, denoted as Fermi gas (FG) in the phase diagram shown in the panel b) of Fig. \ref{Spectral}, corresponds to $\Delta=0$, so that the left and right movers are free particles, decoupled from each other (see inset of Fig.~\ref{fig:figure3}a)). 
In the crystalline phase we have a constant finite value of $\Delta(x)=\Delta_{\rm c}$, with $|\Delta_{\rm c}|=\Delta_0$. The field $\Delta$ thus plays the role of the order parameter for the phase transition. Eq. (\ref{Hamil}) possesses a $U(1)$ translational symmetry which is preserved in the low-energy theory as invariance under the transformation $\Delta\to\Delta \exp(i\chi)$, $u\to u \exp(i\chi/2)$, $v\to v \exp(-i\chi/2)$. This phase does in fact contribute to the coherent part of the full electromagnetic field $a(x)$ through the spatial profile of the emergent crystalline lattice: $\Delta(x)\cos(2k_{\rm F} x+\chi)$, so that $\chi$ fixes the position of the minima of the optical potential.
The additional $\mathbf{Z}_{2}$ symmetry corresponds to the invariance under the transformation $\Psi\to\Psi^{\dagger}$, $\Delta\to -\Delta$. The transition breaks the $U(1)$ symmetry by fixing the phase $\chi$. Here we can choose it such that $\Delta_{\rm c}$ is real. The additional $\mathbf{Z}_2$ symmetry is then broken by choosing the sign of $\Delta_{\rm c}$, which in turn determines the lattice dimerization.
For finite $\Delta_0$, our system now possesses a discrete translational symmetry with spatial period $2\pi/2k_f=\pi/k_{\rm F}$ and is a band insulator where the lower of the two bands: $\epsilon_{k,\sigma}=\pm\sqrt{(v_{\rm F} k)^2+\Delta_0^2}$ is filled (we set the zero of energy at the Fermi energy, $\mu=0$). The inset in Fig.~\ref{fig:figure3}b) shows the gap equal to $2\Delta_0$ that opens in the fermion-spectrum upon entering the crystalline phase.

The low-energy theory of Eq. (\ref{eq:lowenergytheory}) we obtained for our light-matter system turns out to be the same as the continuum limit of the SSH model \cite{ssh_1979,maki_1980} for electrons coupled to lattice vibrations. In our case the role of the lattice phonon is played by the waveguide photon. In particular, given a fixed lattice configuration, the single particle spectrum of the SSH Hamiltonian in the tight-binding limit is determined by the ratio of the two hopping rates defining the dimerization. In our case, the hopping ratio is proportional to the amplitude of the field $\Delta_0$ i.e. to the number of photons in the waveguide with momentum equal to $\pm 2k_{\rm F}$, while which one of the two hoppings is the strongest depends on the sign of $\Delta_{\rm c}$. The latter can be measured from the relative phase ($0$ or $\pi$) between the driving laser and the waveguide output light at momentum $\pm 2k_{\rm F}$.

We will first consider the mean-field solutions given by the saddle-point of the action in Eq. (\ref{eq:lowenergytheory}), which satisfy a set of Bogoliubov-deGennes (BdG) equations for the fields $u,v,\Delta$. 
\begin{align}
\epsilon_\ell \Psi_\ell (x)&=\begin{pmatrix}
-iv_{\rm F}\frac{\partial}{\partial x}&\Delta(x)\\
\Delta^*(x)&iv_{\rm F}\frac{\partial}{\partial x}
\end{pmatrix}{\Psi}_\ell (x)\label{eq:bdgf}\\
\tilde{\delta}_{2k_{\rm F}}\Delta(x)&=-\sum_\ell {'}u_\ell(x)v_\ell^*(x) \label{eq:bdgd}
\end{align}
where we restricted to zero temperature. The index $\ell$ labels the atomic eigenstates $\Psi_\ell^T=(u_\ell,v_\ell)$ and the primed sum runs over the occupied states.
Physically, the BdG equations correspond to neglecting the quantum and thermal fluctuations of the electromagnetic field, still including the fluctuations of the fermions.
Eqs.~(\ref{eq:bdgf}),(\ref{eq:bdgd}) are well studied in the context of the SSH-model. We are thus able to use the solutions obtained there \cite{ssh_1979,maki_1980,brazovskii1980self}, which are presented below (see Supplementary Notes 4 and 5 for more detail).

From the BdG equations we can also obtain approximate analytical expressions for the critical temperature $T_{\rm c}$, as well as the temperature dependence of $\lambda_{\rm c} (2k_{\rm F})$ and the value of $\Delta_0$. 
The critical temperature reads (reinstalling dimensional units)
$
T_{\rm c}\simeq E_{\rm F}\exp(-2\delta_{\rm A}^2 \delta_{2k_{\rm F}}/g^2L\Omega^2 \nu_{k_{\rm F}})),
$ where $\nu_{k_{\rm F}}$ is the fermionic density of states at the Fermi energy, while for the order parameter we get
$
\Delta_0(T)\simeq( 3 T_{\rm c}\delta_{\rm A}/g\Omega\sqrt{L})\sqrt{(T_{\rm c}-T)/T_{\rm c}}.
$
Finally, the critical coupling is given by
\begin{equation}
\lambda_{\rm c}^2\simeq - \frac{\delta_{2k_{\rm F}}}{2\nu_{k_{\rm F}}}\frac{n}{\ln(\frac{E_{\rm F}}{T})}.
\end{equation}

\textbf{Topological soliton-polaritons.} 
Let us now consider the spatially-ordered insulating phase with a filled lower band and study the low-lying excitations. Naturally, there are particle-hole excitations at the energy cost of $2\Delta_0$ which leave the optical field unchanged. However, our system additionally features polaritonic excitations with even lower energy involving a distortion of the optical field together with the excitation of a fermion. Those are found as solutions of the BdG equations whose form is available analytically.
The solitonic distortion of the lattice takes the following form (reinstalling dimensional units)
\begin{equation}
\label{eq:soliton}
\Delta(x,T)=\pm i\Delta_0(T) \tanh(\frac{\lambda\Delta_0(T)}{\sqrt{n}\hbar v_{\rm F}} x).
\end{equation}
The soliton solution of Eq. (\ref{eq:soliton}) (see Fig.~\ref{fig:setup} for the corresponding total light intensity profile) connects a region with negative(positive) $\Delta_0$ at $x\to-\infty$ with a positive(negative) $\Delta_0$ at $x\to+\infty$, that is, it matches two differently dimerized configurations in either one of the two possible ways. Therefore, we can assign to each solitonic defect a $\mathbf{Z}_2$ topological number.
The presence of the solitonic distortion of the lattice creates a single-particle bound state (with spin degeneracy equal to two) which is occupied by a single fermion.  
The full single-particle spectrum is shown in the inset of Fig.~\ref{fig:figure3} $c$). We see that the bound state lies in the middle of the gap between the valence and conduction band. Those bands consist of delocalized states similar to the ones we have in the absence of the defect.
Using the expressions for the critical temperature and the order parameter with Eq. (\ref{eq:soliton}), the size of the optical soliton at zero temperature can be expressed as
\begin{equation}
l_{s0}\sim\frac{2}{3k_{\rm F}}\exp(\frac{2\delta_{\rm A}^2\delta_{2k_{\rm F}}}{g^2\Omega^2 \nu_{k_{\rm F}}})=\frac{2}{3k_{\rm F}}\frac{E_{\rm F}}{T_{\rm c}}.
\end{equation}
In the context of optics, it is interesting to note that the optical soliton's size is set by a scale belonging to the quantum degenerate atoms, namely the inverse Fermi momentum. Moreover, in contrast to their electron-phonon counterparts, the size of the topological soliton-polaritons in our case can be easily tuned by optical parameters, such as laser detunings or pump strength. 

At this point, a remark on the creation of the topological soliton-polaritons is in order. Any local perturbation (adding particles or energy) would necessary create soliton-anti-soliton pairs, since the creation of a single (anti)soliton would require a non-local perturbation. In the context of solid state, topological solitons have been experimentally created by adding several electrons to the dimerized lattice \cite{fournier,monceau}. In this case, every electron seeds the formation of an (anti)soliton, while the odd electron is either absorbed at the boundaries or goes into the conduction band. The scenario is slightly different if the local perturbation adds energy and not particles. In this case, an electron-hole pair would generically be formed and subsequently develop into a soliton and anti-soliton pair. Differently from the case of particle addition where the topological defects are (meta)stable, if only energy is added the defects have a finite lifetime as the soliton will eventually annihilate with the anti-soliton.
The topological protection of the soliton-polaritons relies on the particle-hole symmetry of the crystalline phase. 
It is important to note that strictly speaking this exists only if the crystal does not contain Goldstone modes i.e. phonons that allow for a spatial modulation of the standing-wave phase $\chi$ defined in the subsection ``Low-energy theory''. These modes namely produce modulations of the lattice period which could reabsorb the solitonic defect by accomodating the electron in the band \cite{fournier,monceau}. In the specific configuration considered in Fig.~\ref{fig:setup} there is no externally imposed lattice for the fermions and indeed the crystallization we discuss above breaks a continuous translational invariance, which introduces Goldstone modes. However, due to the typically very long range of the interactions induced by light (the optical waveguide dispersion is indeed much steeper than the fermionic dispersion), the Goldstone-phonon branch will be characterized by a very steep slope or even be gapped if the range is larger than the system's size \cite{ostermann}. This means that over a finite range of energies and times the topological protection would be effectively present. Alternatively, the latter could be exactly implemented by externally imposing an optical lattice potential for the fermions using two far-off detuned counterpropagating lasers. Indeed, in this case the continuous translational $U(1)$-symmetry discussed above will be substituted by a discrete $\textbf{Z}_2$-symmetry, and the Peierls instability would happen at commensurate filling, thereby creating a crystalline insulator without Goldstone phonons.

\textbf{Optical response.} For the setup considered here, there is the possiblity to non-destructively probe the state of the system using the waveguide modes. We shall see that the phase of the system and even the presence of the topological soliton-polaritons can be detected this way via the optical absorption, which is the imaginary part of the dielectric function (see Supplementary Note 6)
\begin{equation}
\Pi(\omega,x,x')=\sum_{\ell\ell'}\frac{v_\ell(x) v_\ell^*(x') u_{\ell'}(x') u_{\ell'}^*(x)}{-\omega+\epsilon_{\ell'}-\epsilon_\ell- i0^+}.
\end{equation}

This function describes how the photons are influenced by interactions with the atomic medium, which involves all possible photon-induced transitions between the single-particle eigenstates of the BdG equation (\ref{eq:bdgf}). By taking the imaginary part we select only resonant processes which indeed are the only ones giving rise to absorption.
In Fig. \ref{fig:figure3} we show the frequency- and momentum-resolved absorption in the spatially homogeneous phase a), in the crystalline phase b), and also in the presence of a soliton-polariton c). In a) and b) the low-energy theory is translational invariant and we define $\mathrm{Im}\Pi(\omega,Q)=\mathrm{Im}\ \int d(x-x')\Pi(\omega,x-x')\exp[i Q (x-x')]$. 
In c), due to the broken translational invariance, $\Pi(\omega,x,x')$ depends upon two positions and we define $\mathrm{Im}\Pi(\omega,Q,Q')=\mathrm{Im}\ \int dx dx'\Pi(\omega,x,x')\exp[i Qx-Q'x')]$. In order to compare to the other two cases we thus consider only the diagonal part $Q=Q'$.

In the homogeneous phase, the optical absorption in Fig.~\ref{fig:figure3} is reminiscent of the spectral function shown in Fig. \ref{Spectral}a) around $Q=2k_{\rm F}$ (recall that we are computing $\Pi(\omega;Q)$ using the low-energy theory where the optical field is expanded about $2k_{\rm F}$).
In the crystalline phase without defects, optical absorption takes place only for probe frequencies $\omega\geq 2\Delta_0$, as resonant excitations into the conduction band are gapped. This is because the minimum difference in energy between the fermion dispersion in the two bands $\epsilon_{k,\sigma}=\pm\sqrt{(v_{\rm F} k)^2+\Delta_0^2}$ is $2\Delta_0$. 
On the other hand, when a soliton-polariton is present in the crystalline phase, the threshold for optical absorption is reduced to $\omega\geq\Delta_0$, due to the appearance of a bound state in the middle of the gap.
The fact that in the vicinity of $\omega\geq\Delta_0$ the main contribution comes from transitions between the bound state and the continuum is also responsible for the flattened bottom of the spectral onset, since the bound state has not a well defined momentum. 
We therefore propose that it is possible to detect the soliton-polaritons nondestructively using the waveguide modes and that their existence should be indicated by optical absorption around $2k_{\rm F}$ at frequencies half the insulating gap.\newline

\textbf{Discussion}

We predicted that the low-lying excitations of a (quasi-)1D Fermi gas coupled to the electromagnetic modes of an optical waveguide are topological soliton-polaritons consisting of an optical soliton trapping an atom in an interphase state localized between two spatially-ordered patterns of different dimerization. While they are formally equivalent to objects predicted for electron-phonon models -- the present system is indeed also an ideal candidate for the implementation of SSH-like physics -- such soliton-polaritons are novel in the context of quantum nonlinear optics, where the topological protection of the polariton might find applications for instance in quantum information storage. In this spirit, future studies should concretely investigate the way such polaritons can be produced in these setups, as well as the role of mutual interactions, quantum fluctuations and dissipation.
Experiments interfacing waveguides with atomic gases are constantly improving the control over the photon dispersion, atomic cooling and light-matter interaction.
Developments are ongoing in several setups like tapered optical nanofibers \cite{rauschenb_2010,rausch}, photonic bandgap fibers \cite{Bandgapfiber} and hollow-core fibers\cite{Hollowcore,PhysRevA.78.033429,bajcsy2011laser}, in addition to hollow-core antiresonant reflecting optics waveguides \cite{arrows} and photonic crystals \cite{Photoniccrystal}. In particular, already some years ago guided light modes have been coupled to quantum degenerate bosons \cite{PhysRevA.78.033429}, making the coupling to fermionic atoms seem a near-future possibility.\\
\textbf{Methods}\newline
\textbf{Effective Action Approach.} Using the imaginary-time path-integral approach as in \cite{PIAZZA2013135,piazza_fermi_2014} we constructed an action from the Hamiltonian in Eq. (\ref{Hamil}) in the main text. We integrated out the atomic degrees of freedom to obtain an effective action for the photons only. The photonic field was then separated into a coherent part and a ``fluctuations" part. Expanding to second order in the fluctuations, we obtained the spectral function and determine the critical point for crystallization. We then derived a low-energy theory for the system by linearizing the atomic dispersion about the Fermi points. The corresponding action is seen in Eq. (\ref{eq:lowenergytheory}) in the main text. The BdG equations for the system, found by calculating the saddle point of the action, are analytically solvable as in \cite{ssh_1979,maki_1980,brazovskii1980self,rice_mele_1982}. As before, we separated into coherent and fluctuation part of the optical field and expanded to second order in fluctuations to obtain the optical response.\\
\textbf{Data availability}
All data generated or analyzed during this study are included in this published article (and its Supplementary Information).
\textbf{Author Contributions}\newline
K.F. constructed the theory, carried out the analytical and numerical calculations, and wrote the paper. F.P. initiated the problem, constructed the theory, and wrote the paper.
\\
\textbf{Additional Information}\\
\textbf{Competing Interests:} The authors declare no competing interests.

\widetext

\begin{center}
\textbf{Supplementary Note 1}
\end{center}

In order to derive the photon-only effective action, we follow the steps delineated in \cite{PIAZZA2013135,piazza_fermi_2014}, we adopt a formulation of the problem based on a path-integral on the imaginary time-axis $\tau\in[0,\beta]$ with $\beta$ the inverse temperature. The action corresponding to the Hamiltonian in Eq. (1)  is quadratic in the atomic fields which can thus be integrated out exactly.
The resulting effective action for the photons, decomposing the field into Matsubara components $a_{k}(\tau)=\frac{1}{\beta}\sum_{n}a_{n,k}e^{-i\omega_n \tau}$ and separating the coherent part $a_{k,0}=\delta_{n,0}^{(\rm K)}\alpha_{k}$ from the fluctuations $a_{k,n}$ ($\delta^{(\rm K)}$ indicates the Kronecker Delta), becomes
\begin{Cequation}
S_{\textrm{eff}}[a_{k}^{\dagger},a_{k}]=\frac{1}{\beta}\sum_{n,k}(-i\omega_n-\delta_{k})|\delta_{n,0}^{(\rm K)}\alpha_{k}+a_{k,n}\theta(|n|)\
|^2-\textrm{Tr}\ln [\textrm{M}],
\end{Cequation}
where the Heaviside step function $\theta(|n|)$ guarantees that the fluctuations are orthogonal to the cohrerent portion. Note that $\omega_n=2\pi nT$. The symbol $\Tr$ denotes a spatial integral and sum in Matsubara space.
\newline
If one denotes
\begin{Cequation}
V_{\textrm{sp}}(x)=\hat{V}(x)\bigg\rvert_{\hat{a}_{k}=\alpha_{k}},
\end{Cequation}
the matrix M is then defined as:
\begin{Cequation}
M_{n,n'}(x,x')=G^{-1}_{n,n'}(x,x')+A_{n,n'}(x,x'),
\end{Cequation} 
with the inverse propagator for the atomic degrees of freedom expressed
\newline
\begin{Cequation}
\begin{split}
G^{-1}_{n,n'}(x,x') & =\beta\biggr[-i\omega_n-\frac{1}{2m}\frac{\mathrm{d}^2}{\mathrm{d}x^2}+V_{\textrm{sp}}(x)-\mu\biggr]\delta_{n,n'}\delta(x-x')
\end{split}
\end{Cequation}
\newline
and the electromagnetic field fluctuations given as the matrix

\begin{Cequation}
\begin{split}
A_{n,n'}(x-x')=& \delta(x-x')\biggr[\sum_{k}\frac{\partial V_{\textrm{sp}}(x)}{\partial \alpha^{}_{k}}a_{k,n-n'}\theta(|n-n'|)+\sum_{k}\frac{\partial V_{\textrm{sp}}(x)}{\partial \alpha_{k}^*}a^{*}_{k,n'-n}\theta(|n'-n|) \\
& +\sum_{k}\frac{\partial^2 V_{\textrm{sp}}(x)}{\partial \alpha^{*}_{k}\partial \alpha_{k'}}a^{*}_{k,n_1+n-n'}a_{k',n_1}\theta(|n_1+n'-n|)\biggr].
\end{split}
\end{Cequation}
\newline
By virtue of the above decomposition of the electromagnetic field into a coherent part and fluctuations, one can separate the effective action into a mean-field (MF) and a fluctuation part:
\begin{Cequation}
S_{\textrm{eff}}[a_{k}^{*},a_{k}]=S^{(MF)}_{\textrm{eff}}+S^{(FL)}_{\textrm{eff}}[a_{k}^{*},a_{k}].
\end{Cequation}

\begin{center}
\textbf{Supplementary Note 2}
\end{center}

In order to compute the optical response in the homogeneous phase, we expand the tracelog to second order in the light-field fluctuations as
$\Tr\ln[M]\approx \Tr\ln[G^{-1}]+\Tr[GA]-\frac{1}{2}\Tr[(GA)^2]$. The second-order effective action in the homogeneous phase ($\alpha=0$) can then be expressed as:
\begin{Cequation}
\label{eq:eff_act_hom}
\begin{split}
S_{\textrm{eff}}^{(2)}[a^*,a]&=-\sum_{Q,\nu}\bigg[\ln[G_0^{-1}(\omega_{\nu};Q)]\\
&+\frac{1}{2\beta}\bigg(a^*_{-Q,\nu} \hspace{1.5mm} a_{Q,-\nu}\bigg)  \begin{pmatrix} -i\omega_{\nu}+\omega_{Q}-\omega_L  -\frac{\lambda^2}{n}\Pi(\omega_{\nu};Q) & -\frac{\lambda^2}{n}\Pi(\omega_{\nu};Q) \\ -\frac{\lambda^2}{n}\Pi(\omega_{\nu};Q) & i\omega_{\nu}+\omega_{Q}-\omega_L -\frac{\lambda^2}{n}\Pi(\omega_{\nu};Q) \end{pmatrix} 
  \begin{pmatrix}a_{-Q,\nu} \\ a^*_{Q,-\nu} \\ \end{pmatrix}\bigg],
\end{split}
\end{Cequation}
where the waveguide's photonic dispersion is denoted $\omega_Q$, the pump frequency is $\omega_L$, and $a^*_{Q,-\nu}$ represents a fluctuation in the optical mean field with momentum $Q$ and Matsubara frequency $\omega_{-\nu}$.

The optical response is characterized by the dielectric (Lindhard) function which in the homogeneous phase can be written $\Pi_F(\omega_{\nu};Q)=-\frac{\beta}{L}\sum_{k,\omega_n}G_0(\omega_n;k)G_0(\omega_{n+\nu};k+Q)$. For the homogeneous phase, we plot in Fig. 2 in the main text the spectral function, defined
\begin{Cequation}
A(\omega;Q)=2\mathcal{I}m[g(-i\omega+0^+;Q)]_{11}=\frac{2(\omega-\delta_Q)^2\mathcal{I}m[\frac{\lambda^2}{n}\Pi(-i\omega+0^+;Q)]}{(\delta_Q^2-\omega^2+2\delta_Q \mathcal{R}e[\frac{\lambda^2}{n}\Pi(-i\omega+0^+;Q)])^2+(2\delta_Q\mathcal{I}m[\frac{\lambda^2}{n}\Pi(-i\omega+0^+;Q)])^2},
\end{Cequation}
with the fluctuation inverse propagator $g^{-1}(\omega_n,Q)$ appearing as the matrix in the second line of eq.~(\ref{eq:eff_act_hom}).
$g$ is also used to determine the condition for crystallization: the latter corresponds to the appearance of zero-frequency modes. One must thus compute the value of $\lambda$ for which $\mathrm{det}[g^{-1}(0,Q)]=0$. In so doing one arrives at the formula quoted in the main text: $\lambda_{SO}^2=-\frac{n\delta_Q}{2\Pi_F(0;Q)}$.
\newline

\begin{center}
\textbf{Supplementary Note 3}
\end{center}

Here we derive the low-energy theory describing our problem.
The Hamiltonian given in Eq.(1) in the main text can be represented in momentum space as 
\begin{Cequation}
H=\sum_{k}\epsilon_k \hat{c}^{\dagger}_k\hat{c}_k+\frac{1}{L}\int\textrm{d}x\sum_{k,k'} \hat{V}(x)e^{i(k-k')x}\hat{c}^{\dagger}_k\hat{c}_{k'}.
\end{Cequation}
with $\hat{V}(x)$ the potential quoted in the main text.
In order to derive a low-energy theory, we linearize the Hamiltonian about the Fermi points and consider scattering processes which transfer either a very small momentum $q\approx0$ or which transfer $q\approx\pm2k_F$, i.e. we linearize also the transferred momenta about $0$ and $\pm 2k_F$. The latter process is known as Umklapp scattering. 
Our Hamiltonian takes thus the form
\begin{Cequation}
\begin{split}
H&= \int \textrm{d}x \bar{\Psi}^{\dagger}(x)\bigg(-iv_F\frac{\partial}{\partial x}\bigg)\sigma_3\bar{\Psi}(x)+ 2\frac{\lambda}{\sqrt{N}} \int \textrm{d}x \bar{\Psi}^{\dagger}(x)\begin{pmatrix}
0 & \Delta(x)\\ \Delta^*(x) & 0
\end{pmatrix}\bar{\Psi}(x)-\int\textrm{d}x {\delta}_{2k_F}|\Delta(x)|^2
\end{split}
\end{Cequation}
From this Hamiltonian we can construct the low-energy action written in Eq. (3) in the main text.
The corresponding BdG equations, given in Eq. (4) of the main text for $T=0$, are obtained from the saddle-point condition. The finite-temperature version of the BdG equations differs from $T=0$ in the self-consistency equation, reading $\tilde{\delta}_{2k_F}\Delta(x)=\sum_{n} (1-2n_F(\epsilon_n^{(-)}))v^*_n(x)u_n(x)$ where $\epsilon_n^{(-)}$ denote the eigenvalues below the chemical potential.

Restricting to the homogeneous phase first where $\Delta(x)= 0$, using the BdG we can calculate the critical temperature  or equivalently the critical coupling strength  (shown in the inset of Fig. 2) of the main text) in the standard way. In the homogeneous phase, the positive and negative branches of the dispersion relation must be considered separately and the BdG equations (see main text) decouple. The resulting solutions are simply plane waves with dispersion relation $\epsilon_k=\pm v_F k$.\\

\begin{center}
\textbf{Supplementary Note 4}
\end{center}

For the crystalline phase, we now calculate the solutions to the BdG equations in the main text for finite and spatially constant $\Delta$. Let us make the variable transformation $\Psi^{i}_l=\frac{1}{\sqrt{2}}\big(\phi^{(1)}_l +(-1)^{i+1} \phi^{(2)}_l\big)$, where $i=1,2$ and decompose $\Delta(x)=\Delta_1(x)+i\Delta_2(x)$ with $|\Delta|=\Delta_0$. 
We are left with:
\begin{Cequation}
\begin{split}
& v_F\partial_x \phi^{(1)}_l-\Delta_2\phi^{(1)}_l-i(E_l+\Delta_1)\phi^{(2)}_l=0,\\
&v_F\partial_x \phi^{(2)}_l+\Delta_2\phi^{(2)}_l-i(E_l-\Delta_1)\phi^{(1)}_l=0.
\end{split}
\end{Cequation}
We next introduce $0\leq \theta\leq \pi$ and $D_0\geq 0$ such that $\Delta_1=\Delta_0\cos\theta$ and
$
D_0=\Delta_0\sin\theta=\sqrt{\Delta_0^2-\Delta_1^2}$.
Following Brazovskii \cite{brazovskii1980self}, let us make a choice of the phase of $\Delta$ such that all of the spatial dependence of $\Delta$ is contained in the imaginary part, i.e $\Delta_1(x)=\Delta_1$ is constant and in the undistorted system the phase of $\Delta$ can be either $0$ or $\pi$. When the distortion is present, it interpolates between the two dimerizations.
In the case of finite (constant) $\Delta$,
we have for the negative-energy (recall that the chemical potential, $\mu=0$) branch
\begin{Cequation}
\label{eq:eigenvectors_dimerized_neg}
\begin{split}
u_k^-(x)=\frac{1}{\sqrt{2N_kL}}\bigg(1-\frac{k}{\epsilon_k+\Delta_0}\bigg)e^{ikx},
v_k^-(x)=\frac{-1}{\sqrt{2N_kL}}\bigg(1+\frac{k}{\epsilon_k+\Delta_0}\bigg)e^{ikx},
\end{split}
\end{Cequation}
while for the positive-energy branch we have
\begin{Cequation}
\label{eq:eigenvectors_dimerized_pos}
\begin{split}
u_k^+(x)=\frac{1}{\sqrt{2M_kL}}\bigg(1+\frac{k}{\epsilon_k-\Delta_0}\bigg)e^{ikx},
v_k^+(x)=\frac{1}{\sqrt{2M_kL}}\bigg(\frac{k}{\epsilon_k-\Delta_0}-1\bigg)e^{ikx}.
\end{split}
\end{Cequation}
The normalisation factors are $N_k=\frac{2\epsilon_k}{\epsilon_k+\Delta_0}$ $M_k=\frac{2\epsilon_k}{\epsilon_k-\Delta_0}$.\\

We note that $\Delta_0$ can be computed by using the critical condition and expanding to leading order in $T-T_c$.

\begin{center}
\textbf{Supplementary Note 5}
\end{center}

In presence of a solitonic defect ($\Delta$ not constant in space), we have (after undoing the variable transformation above)
\begin{Cequation}
\label{eq:eigenvectros_defect}
\begin{split}
&u_0(x)=\sqrt{\frac{D_0}{2v_F}}\sech(\frac{D_0}{v_F}x), v_0(x)=u_0(x),\\
&u_k^-(x)= \frac{e^{ikx}}{\sqrt{2N_kL}}\bigg(\frac{-v_Fk+iD_0\mathrm{tanh}(\frac{D_0}{v_F}x)}{\epsilon_k}+1\bigg) , v_k^-(x)=\frac{e^{ikx}}{\sqrt{2N_kL}}\bigg(\frac{-v_Fk+iD_0\mathrm{tanh}(\frac{D_0}{v_F}x)}{\epsilon_k}-1\bigg),\\
&u_k^+(x)= \frac{e^{ikx}}{\sqrt{2N_kL}}\bigg(\frac{v_Fk-iD_0\mathrm{tanh}(\frac{D_0}{v_F}x)}{\epsilon_k}+1\bigg) , v_k^+(x)=\frac{e^{ikx}}{\sqrt{2N_kL}}\bigg(\frac{v_Fk-iD_0\mathrm{tanh}(\frac{D_0}{v_F}x)}{\epsilon_k}-1\bigg)
\end{split}
\end{Cequation}
for the atomic eigenfunctions in the bound, negative-energy and positive-energy states, respectively, where the normalisation constant is the same for both the upper and lower bands and is now given by $
N_k= 2$.
For some value of $l$, let us say $l=0$, there exists a localised solution.
If one substitutes the first of the BdG equations into the other, the limiting behaviour of the resulting second-order differential equation are as follows: $x\to\pm\infty\implies\Delta'_2\to 0$, $\Delta^2_2+E_0^2\to\Delta_0^2$ and therefore
\begin{Cequation}
v_F^2\phi^{(2)''}_l+(E^2_l-\Delta_0^2)\phi^{(2)}_l=0.
\end{Cequation}
The solutions to this equation are plane waves $\phi^{(2)}_k(x)=\frac{1}{\sqrt{N_k L}}\exp(ikx)$ with eigenvalues $E_k=\epsilon_k=\pm\sqrt{(v_Fk)^2+\Delta_0^2}$, where $L$ denotes the sample length, $N_k$ is a normalisation factor and we denoted the undeformed solutions with the index $l=k$.
Now, these plane waves must satisfy the full differential equation. Upon substituting, the result is
\begin{Cequation}
\begin{split}
-v_F\frac{\mathrm{d}\Delta_2}{\mathrm{d}x}+\Delta^2_2=D_0^2
\end{split}
\end{Cequation}
which is solved by 
\begin{Cequation}
\Delta_2(x)=-D_0\tanh(\frac{D_0}{v_F} x).
\end{Cequation}
The corresponding solution to the BdG equations is given by
$E_{0}=\Delta_1$, $\phi^{(2)}_{0}=0$ and 
\begin{Cequation}
\phi^{(1)}_{0}(x)=\sqrt{\frac{D_0}{2v_F}}\sech(\frac{D_0}{v_F} x)
\end{Cequation}
 and the saddle-point equations of the action (see main text) are:
\begin{Cequation}
\tilde{\delta}_{2k_F}\Delta_1=\sum_l \phi^{(2)*}_l\phi^{(2)}_l-\phi^{(1)*}_l\phi^{(1)}_l, \hspace{1cm}
\tilde{\delta}_{2k_F}\Delta_2=i\sum_l \phi^{(1)*}_l\phi^{(2)}_l-\phi^{(2)*}_l\phi^{(1)}_l.
\end{Cequation}

We can now calculate the elastic energy of the lattice to determine the stability of the defect state. If one denotes the spin degeneracy of the bands by $\nu$ and the occupation of the bound state by $\nu_0$, one obtains
\begin{Cequation}
\begin{split}
\frac{\mathrm{d}E(\theta)}{\mathrm{d}\theta}=\int\mathrm{d}x\bigg(\frac{\delta S}{\delta\Delta_1}\frac{\mathrm{d}\Delta_1}{\mathrm{d}\theta}+\frac{\delta S}{\delta\Delta_2}\frac{\mathrm{d}\Delta_2}{\mathrm{d}\theta}\bigg)
=\Delta_0\sin\theta\bigg(\nu\frac{\theta}{\pi}-\nu_0\bigg).
\end{split}
\end{Cequation}
Integrating:
\begin{Cequation}
E(\theta)=\Delta_0\bigg[\bigg(\nu_0-\nu\frac{\theta}{\pi}\bigg)\cos\theta+\frac{\nu}{\pi}\sin\theta\bigg], E_s=\frac{\nu\Delta_0}{\pi}\sin\bigg(\pi\frac{\nu_0}{\nu}\bigg).
\end{Cequation}
By inspection, one sees that the energy has a stationary point at $\theta=0,\frac{\pi\nu_0}{\nu}$. By calculation of the second derivative of $E(\theta)$, one finds that the energy is maximised for $\theta=0$ and minimised for $\theta=\theta_0\equiv\frac{\pi\nu_0}{\nu}$.
Since $0\leq\nu_0 \leq \nu$, for degeneracy $\nu=0,1$, there are two possibilities:
1. $\nu_0=\theta_0=D_0=0\implies$ Undeformed. 
2. $\nu_0=\nu\implies\theta_0=\pi\implies D_0=0\implies$ Undeformed with $\nu$-fold degenerate ground state.
For $\nu=2$ one has a non-trivial stationary bound state with $\nu_0=1\implies\theta_0=\frac{\pi}{2}\implies D_0=\Delta_0/v_F, E_0=\Delta_1=0, E_s=\frac{2}{\pi}\Delta_0$ and thus the defect is stable.\\

\begin{center}
\textbf{Supplementary Note 6}
\end{center}

In order to compute the optical response, we perform the same analysis for the low-energy theory as we did before the full Hamiltonian in the homogeneous phase. That is, we separate the optical field $\Delta$ into a coherent mean-field part and the fluctuations: 
\begin{Cequation}
\begin{split}
&S=S_{\mathrm{ph}}+S_{\mathrm{a}}+S_{\mathrm{a/p}}:\\
&S_{\mathrm{ph}}=\frac{1}{\beta}\int\mathrm{d}x\sum_n(\Delta_n^*\delta_{n,0}+\delta\Delta_n^*\theta(|n|))(-i\frac{\sqrt{n}}{2\lambda}\omega_n+\tilde{\delta}_{2k_F})(\Delta_n\delta_{n,0}+\delta\Delta_n\theta(|n|));\\
&S_{\mathrm{a}}=\frac{1}{\beta}\int\mathrm{d}x\sum_n\bar{\Psi}^{\dagger}_n(-i\omega_n-iv_F\partial_x\sigma_3)\bar{\Psi}_n;\\
&S_{\mathrm{a/p}}=\frac{1}{\beta^2}\int\mathrm{d}x\sum_{n,m}\bar{\Psi}^{\dagger}_n\sigma_1\mathrm{diag}(\Delta^*_m\delta_{m-n,0}+\delta\Delta^*_{m-n}\theta(|m-n|),\Delta_{n-m}\delta_{n-m,0}+\delta\Delta_{n-m}\theta(|n-m|))\bar{\Psi}_{m},
\end{split}
\end{Cequation}
where again $\omega_n$ denotes a Matsubara frequency, and expand to second order in the fluctuations.\\

We characterize the optical response by means of the imaginary part of the dielectric function $\Pi$, as this gives us the amount of absorption from the medium. In Fig. 3 of the main text, we plot the top left entry of $\Pi$, which in this case is a matrix in the same basis (though the fluctuations are now in $\Delta$, not $a$) as the full spectral function $A(\omega,Q)$.

In the homogeneous phase, the functions $v_{\ell}(x)$ and $u_{\ell}(x)$ are plane waves and substituting into Eq.9 in the main text we obtain the optical absorption, which is given by $\frac{1}{4v_F}$ for $Q-2k_F>|\omega|$ and is zero otherwise (see Fig. 3 a) in the main text). 

For the crystalline insulator phase, substituting the solutions (\ref{eq:eigenvectors_dimerized_neg}),~(\ref{eq:eigenvectors_dimerized_pos}) into Eq. 9 in the main text, we separate the Greens function into the sum of a negative- and positive-energy branch part. We find that the possible transitions at zero temperature are only those from the negative branch to the positive branch. Evaluating one of the Matsubara sums with $m=\nu-n$, after analytic continuation ($\omega_n\to -i\omega+0^+$) we arrive at the following for the optical absorption plotted in Fig. 3(b) of the main text:
\begin{Cequation}
\begin{split}
&\sum_{k}
\frac{\pi}{4LN_k M_{k+Q}}
\delta(-\omega+\epsilon^{(+)}_{k+Q}-\epsilon^{(-)}_k)
\bigg(1-\frac{k}{\epsilon_k+\Delta_0}\bigg)^2\bigg(\frac{k+Q}{\epsilon_k-\Delta_0}-1\bigg)^2\\
\end{split}
\end{Cequation}
In the case of the crystalline insulator with the defect, we repeat the same calculation this time substituting the solutions in (\ref{eq:eigenvectros_defect}) into Eq. 9 of the main text. Considering the bound state, positive energy, and negative energy branches, we separate the Greens function into the sum of several contributions.  In this case, the possible transitions at zero temperature are those from the negative branch to the positive branch, the negative branch to the bound state at zero energy and from the occupied bound state (say for example the $\downarrow$ state) to the upper branch. The corresponding term for the plot in Fig. 3(c) of the main text is composed of three parts:

The lower to upper branch scattering is represented by ($m=\nu-n$)
\begin{Cequation}
\begin{split}
&\int\mathrm{d}x\int\mathrm{d}x'\sum_{n,\nu,k,k'}\frac{u^{(-)}_k(x)u^{(-)*}_k(x')}{-i\omega_n+\epsilon^{(-)}_k}\frac{v^{(+)}_{k'}(x')v^{(+)*}_{k'}(x)}{-i\omega_{\nu}+\epsilon^{(+)}_{k'}}\delta\Delta_{n-\nu}^*(x)\delta\Delta_{n-\nu}(x')\\
&=\sum_{m,k,k'} \frac{1}{4L^2N_kM_{k'}}\frac{n_F(\epsilon^{(-)}_k)-n_F(\epsilon^{(+)}_{k'})}{-i\omega_m+\epsilon^{(+)}_{k'}-\epsilon^{(-)}_k}\int\mathrm{d}x\int\mathrm{d}x'\delta\Delta^*_{-m}(x)\delta\Delta_{-m}(x')e^{i(k-k')(x-x')}\\
&\bigg[\bigg(1-\frac{v_Fk}{\epsilon_k}\bigg)^2\bigg(1-\frac{v_Fk'}{\epsilon_{k'}}\bigg)^2+\frac{D_0^4}{\epsilon^2_k\epsilon^2_{k'}}\tanh(\frac{D_0}{v_F}x)^2\tanh(\frac{D_0}{v_F}x')^2\\
&+\frac{D_0^2}{\epsilon^2_{k'}}\bigg(1-\frac{v_Fk}{\epsilon_k}\bigg)^2\tanh(\frac{D_0}{v_F}x)\tanh(\frac{D_0}{v_F}x')+\frac{D_0^2}{\epsilon^2_k}\bigg(1-\frac{v_Fk'}{\epsilon_{k'}}\bigg)^2\tanh(\frac{D_0}{v_F}x)\tanh(\frac{D_0}{v_F}x')\\
&+\frac{D_0^2}{\epsilon_k\epsilon_{k'}}\bigg(1-\frac{v_Fk}{\epsilon_k}\bigg)\bigg(1-\frac{v_Fk'}{\epsilon_{k'}}\bigg)\bigg(\tanh(\frac{D_0}{v_F}x)-\tanh(\frac{D_0}{v_F}x')\bigg)^2\\
&+\frac{iD_0}{\epsilon_{k'}}\bigg(\bigg(1-\frac{v_Fk'}{\epsilon_{k'}}\bigg)^2\bigg(1-\frac{v_Fk}{\epsilon_k}\bigg)-\bigg(1-\frac{v_Fk}{\epsilon_{k}}\bigg)^2\bigg(1-\frac{v_Fk'}{\epsilon_{k'}}\bigg)\bigg)\bigg(\tanh(\frac{D_0}{v_F}x)-\tanh(\frac{D_0}{v_F}x')\bigg)\\
&+\bigg(\frac{iD_0^3}{\epsilon_k\epsilon^2_{k'}}\bigg(1-\frac{v_Fk}{\epsilon_k}\bigg)-\frac{iD_0^3}{\epsilon^2_k\epsilon_{k'}}\bigg(1-\frac{v_Fk'}{\epsilon_{k'}}\bigg)\bigg)\bigg(\tanh(\frac{D_0}{v_F}x)-\tanh(\frac{D_0}{v_F}x')\bigg)\tanh(\frac{D_0}{v_F}x)\tanh(\frac{D_0}{v_F}x')\bigg]
\end{split}
\end{Cequation}
The lower branch to bound state scattering term is calculated thus
\begin{Cequation}
\begin{split}
&\int\mathrm{d}x\int\mathrm{d}x'\sum_{n,\nu,k}\frac{u^{(-)}_{k}(x)u^{(-)*}_{k}(x')}{-i\omega_n+\epsilon^{(-)}_k}
\bigg(\frac{v^{(0)}_{\uparrow}(x')v^{(0)*}_{\uparrow}(x)}{-i\omega_{\nu}+E_0-\mu_{\uparrow}}+\frac{v^{(0)}_{\downarrow}(x')v^{(0)*}_{\downarrow}(x)}{-i\omega_{\nu}+E_0-\mu_{\downarrow}}\bigg)\delta\Delta_{n-\nu}^*(x)\delta\Delta_{n-\nu}(x')\\
&=\sum_{m,k}\frac{D_0}{4v_FLN_k}\frac{n_F(\epsilon_k^{(-)})-n_F(E_0-\mu_{\uparrow})}{-i\omega_m+E_0-\mu_{\uparrow}-\epsilon^{(-)}_k}\int\mathrm{d}x\int\mathrm{d}x' e^{ik(x-x')}\delta\Delta_{-m}(x')\delta\Delta^*_{-m}(x)\sech(\frac{D_0}{v_F}x)\sech(\frac{D_0}{v_F}x')\\
&\bigg(\bigg(1-\frac{v_Fk}{\epsilon_k}\bigg)^2+\frac{iD_0}{\epsilon_k}\bigg(1-\frac{v_Fk}{\epsilon_k}\bigg)\bigg(\tanh(\frac{D_0}{v_F}x)-\tanh(\frac{D_0}{v_F}x')\bigg)+\frac{D_0^2}{\epsilon^2_k}\tanh(\frac{D_0}{v_F}x)\tanh(\frac{D_0}{v_F}x')\bigg),
\end{split}
\end{Cequation}
and finally for the process from the bound state to the upper band, 
\begin{Cequation}
\begin{split}
&\int\mathrm{d}x\int\mathrm{d}x'\sum_{n,\nu,k'}\bigg(\frac{u^{(0)}_{\downarrow}(x)u^{(0)*}_{\downarrow}(x')}{-i\omega_n+E_0-\mu_{\downarrow}}
+\frac{u^{(0)}_{\uparrow}(x)u^{(0)*}_{\uparrow}(x')}{-i\omega_n+E_0-\mu_{\uparrow}}\bigg)\frac{v^{(+)}_{k'}(x')v^{(+)*}_{k'}(x)}{-i\omega_{\nu}+\epsilon^{(+)}_{k'}}\delta\Delta_{n-\nu}^*(x)\delta\Delta_{n-\nu}(x')\\
&=\sum_{m,k'}\frac{D_0}{4v_FLM_{k'}}\frac{n_F(\epsilon_{k'}^{(+)})-n_F(E_0-\mu_{\downarrow})}{-i\omega_m-E_0+\mu_{\downarrow}+\epsilon^{(+)}_{k'}}\int\mathrm{d}x\int\mathrm{d}x' e^{-ik'(x-x')}\delta\Delta_{-m}(x')\delta\Delta_{m}(x)\sech(\frac{D_0}{v_F}x)\sech(\frac{D_0}{v_F}x')\\
&\bigg(\bigg(\frac{v_Fk'}{\epsilon_{k'}}-1\bigg)^2+\frac{iD_0}{\epsilon_{k'}}\bigg(\frac{v_Fk'}{\epsilon_k}-1\bigg)\bigg(\tanh(\frac{D_0}{v_F}x)-\tanh(\frac{D_0}{v_F}x')\bigg)+\frac{D_0^2}{\epsilon^2_k}\tanh(\frac{D_0}{v_F}x)\tanh(\frac{D_0}{v_F}x')\bigg).
\end{split}
\end{Cequation}
Collecting these terms we can construct a Lindhard function dependent on two spatial positions, $\Pi(\omega;x,x')$:
\begin{Cequation}
\begin{split}
\Pi(\omega;x,x')&=\sum_{k,k'} \frac{1}{4L^2N_kM_{k'}}\frac{n_F(\epsilon^{(-)}_k)-n_F(\epsilon^{(+)}_{k'})}{-i\omega_m+\epsilon^{(+)}_{k'}-\epsilon^{(-)}_k}e^{i(k-k')(x-x')}\bigg[\bigg(1-\frac{v_Fk}{\epsilon_k}\bigg)^2\bigg(1-\frac{v_Fk'}{\epsilon_{k'}}\bigg)^2+\frac{D_0^4}{\epsilon^2_k\epsilon^2_{k'}}\tanh(\frac{D_0}{v_F}x)^2\tanh(\frac{D_0}{v_F}x')^2\\
&+\frac{D_0^2}{\epsilon^2_{k'}}\bigg(1-\frac{v_Fk}{\epsilon_k}\bigg)^2\tanh(\frac{D_0}{v_F}x)\tanh(\frac{D_0}{v_F}x')+\frac{D_0^2}{\epsilon^2_k}\bigg(1-\frac{v_Fk'}{\epsilon_{k'}}\bigg)^2\tanh(\frac{D_0}{v_F}x)\tanh(\frac{D_0}{v_F}x')\\
&+\frac{D_0^2}{\epsilon_k\epsilon_{k'}}\bigg(1-\frac{v_Fk}{\epsilon_k}\bigg)\bigg(1-\frac{v_Fk'}{\epsilon_{k'}}\bigg)\bigg(\tanh(\frac{D_0}{v_F}x)-\tanh(\frac{D_0}{v_F}x')\bigg)^2\\
&+\frac{iD_0}{\epsilon_{k'}}\bigg(\bigg(1-\frac{v_Fk'}{\epsilon_{k'}}\bigg)^2\bigg(1-\frac{v_Fk}{\epsilon_k}\bigg)-\bigg(1-\frac{v_Fk}{\epsilon_{k}}\bigg)^2\bigg(1-\frac{v_Fk'}{\epsilon_{k'}}\bigg)\bigg)\bigg(\tanh(\frac{D_0}{v_F}x)-\tanh(\frac{D_0}{v_F}x')\bigg)\\
&+\bigg(\frac{iD_0^3}{\epsilon_k\epsilon^2_{k'}}\bigg(1-\frac{v_Fk}{\epsilon_k}\bigg)-\frac{iD_0^3}{\epsilon^2_k\epsilon_{k'}}\bigg(1-\frac{v_Fk'}{\epsilon_{k'}}\bigg)\bigg)\bigg(\tanh(\frac{D_0}{v_F}x)-\tanh(\frac{D_0}{v_F}x')\bigg)\tanh(\frac{D_0}{v_F}x)\tanh(\frac{D_0}{v_F}x')\bigg]\\
&+\sum_{k}\frac{D_0}{4v_FLN_k}\frac{n_F(\epsilon_k^{(-)})-n_F(E_0-\mu_{\uparrow})}{-i\omega_m+E_0-\mu_{\uparrow}-\epsilon^{(-)}_k} e^{ik(x-x')}\sech(\frac{D_0}{v_F}x)\sech(\frac{D_0}{v_F}x')\\
&\bigg(\bigg(1-\frac{v_Fk}{\epsilon_k}\bigg)^2+\frac{iD_0}{\epsilon_k}\bigg(1-\frac{v_Fk}{\epsilon_k}\bigg)\bigg(\tanh(\frac{D_0}{v_F}x)-\tanh(\frac{D_0}{v_F}x')\bigg)+\frac{D_0^2}{\epsilon^2_k}\tanh(\frac{D_0}{v_F}x)\tanh(\frac{D_0}{v_F}x')\bigg)+\\
&+\sum_{k'}\frac{D_0}{4v_FLM_{k'}}\frac{n_F(\epsilon_{k'}^{(+)})-n_F(E_0-\mu_{\downarrow})}{-i\omega_m-E_0+\mu_{\downarrow}+\epsilon^{(+)}_{k'}}e^{-ik'(x-x')}\sech(\frac{D_0}{v_F}x)\sech(\frac{D_0}{v_F}x')\\
&\bigg(\bigg(\frac{v_Fk'}{\epsilon_{k'}}-1\bigg)^2+\frac{iD_0}{\epsilon_{k'}}\bigg(\frac{v_Fk'}{\epsilon_k}-1\bigg)\bigg(\tanh(\frac{D_0}{v_F}x)-\tanh(\frac{D_0}{v_F}x')\bigg)+\frac{D_0^2}{\epsilon^2_k}\tanh(\frac{D_0}{v_F}x)\tanh(\frac{D_0}{v_F}x')\bigg)
\end{split}
\end{Cequation}

Fourier transforming in the two spatial variables, we obtain a Lindhard function dependent upon two momenta, $\Pi(\omega;q,q')$. We then evaluate this object on the diagonal in momentum space in order to compare it with the other two cases.


%

\end{document}